\documentclass[sigconf]{acmart}
\usepackage{tabularx,multirow,multicol}
\AtBeginDocument{%
  }

\setcopyright{acmlicensed}
\copyrightyear{2018}
\acmYear{2018}
\acmDOI{XXXXXXX.XXXXXXX}
\acmConference[Conference acronym 'XX]{Make sure to enter the correct
  conference title from your rights confirmation email}{June 03--05,
  2018}{Woodstock, NY}
\acmISBN{978-1-4503-XXXX-X/2018/06}

\begin{document}

\title{Understanding Users' Privacy Perceptions Towards LLM's RAG-based Memory}

\author{Shuning Zhang}
\orcid{0000-0002-4145-117X}
\email{zsn23@mails.tsinghua.edu.cn}
\affiliation{%
  \institution{Tsinghua University}
  \city{Beijing}
  \country{China}
}

\author{Rongjun Ma}
\email{rongjun.ma@aalto.fi}
\affiliation{
  \institution{Aalto University}
  \city{Espoo}
  \country{Finland}
}

\author{Ying Ma}
\orcid{0000-0001-5413-0132}
\email{ying.ma1@student.unimelb.edu.au}
\affiliation{%
  \department{School of Computing and \\Information Systems}
  \institution{University of Melbourne}
  \city{Melbourne}
  \country{Australia}}

\author{Shixuan Li}
\orcid{0009-0008-6828-6347}
\email{li-sx24@mails.tsinghua.edu.cn}
\affiliation{
    \institution{Tsinghua University}
    \city{Beijing}
    \country{China}
}

\author{Yiqun Xu}
\orcid{0009-0004-9774-9129}
\email{xuyiqun22@mails.tsinghua.edu.cn}
\affiliation{%
  \institution{Tsinghua University}
  \city{Beijing}
  \country{China}
}

\author{Xin Yi}
\orcid{0000-0001-8041-7962}
\authornote{Corresponding author.}
\email{yixin@tsinghua.edu.cn}
\affiliation{
    \institution{Tsinghua University}
    \city{Beijing}
    \country{China}
}

\author{Hewu Li}
\orcid{0000-0002-6331-6542}
\email{lihewu@cernet.edu.cn}
\affiliation{
    \institution{Tsinghua University}
    \city{Beijing}
    \country{China}
}

\renewcommand{\shortauthors}{Trovato et al.}

\begin{abstract}
  Large Language Models (LLMs) are increasingly integrating memory functionalities to provide personalized and context-aware interactions. However, user understanding, practices and expectations regarding these memory systems are not yet well understood. This paper presents a thematic analysis of semi-structured interviews with 18 users to explore their mental models of LLM's Retrieval Augmented Generation (RAG)-based memory, current usage practices, perceived benefits and drawbacks, privacy concerns and expectations for future memory systems. Our findings reveal diverse and often incomplete mental models of how memory operates. While users appreciate the potential for enhanced personalization and efficiency, significant concerns exist regarding privacy, control and the accuracy of remembered information. Users express a desire for granular control over memory generation, management, usage and updating, including clear mechanisms for reviewing, editing, deleting and categorizing memories, as well as transparent insight into how memories and inferred information are used. We discuss design implications for creating more user-centric, transparent, and trustworthy LLM memory systems.
\end{abstract}




\received{20 February 2007}
\received[revised]{12 March 2009}
\received[accepted]{5 June 2009}

\maketitle

\section{Introduction}
Large Language Models (LLMs) like ChatGPT\footnote{\url{https://chat.openai.com/}}, Gemini\footnote{\url{https://gemini.google.com/}}, and Kimi\footnote{\url{kimi.moonshot.cn}} are rapidly evolving from stateless tools into personalized assistants. A key technology driving this shift is the integration of memory, which allows LLMs to retain information across conversations to provide more coherent and contextually aware interactions. While this promises enhanced utility, it also introduces a significant and complex privacy challenge. By creating a persistent record of user interactions, these systems build detailed profiles that can include sensitive thoughts, personal plans, and confidential information, moving beyond transient queries to continuous user data collection.

This capability creates a fundamental tension between personalization and privacy of LLM's memory. On one hand, users desire the efficiency and tailored responses that memory enables. On the other, the opacity of how these systems remember, analyze and utilize personal data can undermine a user's sense of control and informational self-determination. The ``black box'' nature of LLMs exacerbates this issue, leaving users unable to fully understand the scope of data being collected in memories or the associated privacy risks. Specifically, for RAG-based memories, they were usually extracted from users' dialogues, as per ChatGPT, and used for latter personalized dialogues as context information. It may cause potential risks if those RAG-based memories are utilized for training or disclosed to untrusted third-party services. However, how users understand the RAG-based memories, what's their practice and challenges remains under-explored. This study therefore investigates how users perceive RAG-based memories, manage the conflict between privacy and utility, and what's their challenges. We aim to first understand users' mental models of memories as their mental model largely influence whether they could are aware of the potential privacy risks and could effectively control. We then identified their current calculus around the privacy-utility trade-offs and the protective strategies, as private memories inherently possess the conflicts between personalization and privacy leakage. We finally understand users' challenges and expectations for improving RAG-based memory systems. Specifically, we seek to answer the following questions:

RQ1. What is users' mentral models of RAG-based LLM memories?
% How do users' mental models of LLM memory shape their perceptions of privacy risk and their ability to give informed consent?

RQ2. How do users navigate the trade-offs between utility and privacy when using or considering LLM memory, and what privacy protection strategies do they currently employ?

RQ3. What are the challenges users face when attempting to align an LLM's memory behavior with their privacy goals?

We conducted semi-structured interviews with 18 Chinese participants from diverse backgrounds with varying levels of experience. To anchor the discussion in a real-world context and elicit concrete responses, the interview was centered on ChatGPT's memory feature as a prominent case study. 
% such as engineering, IT, design, and social sciences. Participants reported varying levels of experience with LLMs, from daily use to less frequent interactions. To anchor the discussion in a real-world context and elicit concrete responses, the interview was centered on ChatGPT's memory feature as a prominent case study. 

Towards RQ1, we find four prominent mental models of users: users regard memory as a transient, dialogue-specific buffer, as an extension of the core training data, as an active information processing mechanism, and some users explicitly acknowledged the lack of understanding. These mental models reflected different understandings and cognitive processing logics of LLMs' memories. Towards RQ2, our research shows that users are not passive but are active agents performing a continuous privacy calculus, weighing functional benefits against perceived risks. The primary benefits involved enhance personalization and efficiency, improvec continuity and reduced redundancy, and LLM to be a better companion of assistant. The primary drawbacks include problems such as confidentiality and data leakage, profiling and unwanted persuasion, aggregation and re-identification, unauthorized secondary use, and the lack of epistemological certainty. Users employ proactive protective strategies, such as strategic privacy disclosure, proactive input obfuscation and refusal to use or workarounds. Towards RQ3, we identify an unequivocal user mandate for a fundamental shift toward user-directed memory systems built on granular control and transparency across the entire data lifecycle. This includes explicit consent at the point of generation, comprehensive management interfaces for editing and deletion, purpose limitation controls during usage, and direct agency over system-inferred information. To sum up, our contributions are:

$\bullet$ We characterize users' diverse mental models of LLM memory and their privacy risk perceptions.

$\bullet$ We investigate the privacy calculus and protective strategies users employ, such as data minimization and anonymization.

$\bullet$ We identify the critical privacy challenges user face and distill our findings into design implications for user-centric, privacy-preserving memory systems.

\section{Related Work}

\subsection{Privacy Protection of Text-based LLMs}

Research on privacy protection for LLMs follows two primary lines: privacy risk evaluation and direct defense mechanisms. Risk evaluation, which underpins any protective effort, is supported by several toolkits and benchmarks. LLM-PBE~\cite{li2024llm} supplies a structured framework for assessing risks throughout the model lifecycle, and PrivLM-Bench~\cite{li2024privlm} offers a standardized benchmark to quantify data leakage. Specialized instruments such as PrivacyLens~\cite{shaoprivacylens} evaluate compliance with privacy norms, while ProPILE~\cite{kim2023propile} probes models to detect possible PII exposure.

Direct defense mechanisms are engineered to intervene at different points. Some operate at the data interface: OneShield~\cite{asthana2025deploying} filters both user inputs and model outputs, while Rescriber~\cite{zhou2025rescriber} leverages an LLM to minimize sensitive content in queries in real time. In contrast, other strategies intervene at the model level. The CPPLM paradigm~\cite{xiao2024large}, for example, embeds safeguards directly into the fine-tuning process to protect inference-time privacy.

\subsection{Understanding Users' Privacy Concerns of LLMs}

Research into users' privacy concerns in digital contexts began early, showing that users often self-regulate by selectively withholding information or avoiding services perceived as high-risk~\cite{milne2017information,olson2005study,barbosa2020privacy}. In the LLM domain, this manifests as a persistent trade-off between privacy, utility, and convenience~\cite{zhang2024s}, frequently giving rise to a “privacy paradox” in which users tolerate greater information leakage in exchange for higher utility~\cite{zhang2024privacy}. Similarly, users of conversational agents tend to exhibit fewer privacy concerns than non-users, sharing sensitive lifestyle and health data while withholding direct identifiers~\cite{zufferey2025ai}.

These behaviors are further amplified by cognitive biases and system design. Users often rely on inaccurate mental models of LLM data flows and are vulnerable to dark patterns, which impedes their ability to grasp real privacy risks~\cite{zhang2024s,ma2025privacy}. Additionally, the anthropomorphic presentation of LLMs encourages oversharing, as users overestimate the system’s capabilities and attribute human-like understanding to it~\cite{pairguidebook2019,weidinger2022taxonomy}. This tendency heightens privacy exposure and can be exploited for malicious purposes~\cite{pardes2018emotional,ischen2020privacy}.  

\subsection{Memory in LLM-driven Agents}

The capacity to maintain conversational history is a foundational feature of LLM-driven agents, enabling coherent and contextually aware human–AI interaction~\cite{fan2024survey,liu2023think}. Early methods appended the full chat history to the model context. However, as conversations lengthened, agent performance degraded due to distraction and information loss—often called ``lost in the middle''~\cite{shi2023large,liu2023lost}. Consequently, research shifted toward more sophisticated memory architectures. These include recursive summarization and refinement techniques to distill salient information and reduce redundancy~\cite{wang2023recursively,xu2021beyond,xu2022,zhong2022less,huang2023memory}, as well as selective retrieval-based systems that store memories in unstructured or layered repositories and fetch them based on conversational relevance~\cite{bae2022keep,openai2024memory,wu2023brief,zhao2022improving,yu2024finmem}.

Despite these architectural advances, memory storage and retrieval remain opaque to end users, impeding effective task integration. Recent work has explored interactive memory systems that give users direct control via operational ``sandboxes'' for manual editing~\cite{huang2023memory} or visual interfaces for intuitive organization~\cite{yen2024memolet}. Zhang et al.~\cite{zhang2024ghost} further examined user challenges and developed techniques to support memory use. However, these studies have not conducted an in-depth analysis of users' mental models of memory systems nor explicitly examined the core trade-off between personalization and privacy, which this work addresses.

\section{Methodology}

The study employed a qualitative approach to investigate user perceptions and expectations of memory functionalities in LLMs. We utilized semi-structured interviews to gather rich, in-depth data from participants.

\subsection{Participant Recruitment and Demographics}

We recruited 18 participants for this study through distributing questionnaires online. The participants represented a range of academic and professional backgrounds, including fields such as engineering, life sciences, social sciences, IT and design. Table~\ref{tbl:demographics} showed the demographics of participants. The study was approved by the Institutional Review Board (IRB) of our institution, and each participant was compensated 100 RMB for their time.

\begin{table*}[htbp]
\caption{Demographics of users. (`Par' denotes participant ID.)}
\label{tbl:demographics}
\centering
\begin{tabularx}{\textwidth}{p{0.6cm}p{0.5cm}p{1.7cm}Xp{5.8cm}}
\toprule
\textbf{Par} & \textbf{Age} & \textbf{\shortstack{Highest \\ education}} & \textbf{Usage experience} & \textbf{Occupation / } \\
\midrule
P1 & 23 & Ph.D. & ChatGPT, kimi, openai chat sider, dxyz, mobile poe, slack & Clean combustion \\
P2 & 21 & High school & ChatGPT, Gemini, Kimi & Industrial engineering \\
P3 & 22 & Bachelor & ChatGPT3.5/4o & Naval architecture and ocean engineering \\
P4 & 21 & Bachelor & ChatGPT, ChatGPT4.0 & Engineering mechanics \\
P5 & 24 & Master & ChatGPT, Wenxinyiyan, Kimi, Bing & Other \\
P6 & 20 & Bachelor & ChatGPT & Electronic packaging \\
P7 & 19 & High school & GPT, Kimi & Electronic science and technology, AI research \\
P8 & 24 & Bachelor & Kimi, ChatGPT3.5 & Life science \\
P9 & 22 & Master & ChatGPT3.5, Kimi, Wenxinyiyan & Social science \\
P10 & 26 & Master & ChatGPT, ChatGPT4.0 & Civil engineering, rock mechanics \\
P11 & 24 & Master & ChatGPT, Tongyi, PI, free, no payment & Other \\
P12 & 20 & Bachelor & ChatGPT, Gemini & IT related, software engineering \\
P13 & 24 & Master & ChatGPT, Kimi & Not disclosed \\
P14 & 27 & Master & GPT, Wenxinyiyan, Kimi, Poe, TongYi, Doubao, Xiaomei & Design related \\
P15 & 23 & Master & ChatGPT, ChatGLM, Kimi,  Doubao & Electronic information, digital media \\
P16 & 24 & Master & ChatGPT, Kimi, Doubao & Design studies \\
P17 & 22 & Bachelor & ChatGPT, Midjourney, Stable Diffusion & Design profession \\
P18 & 24 & Master & ChatGPT & Software engineering \\
\bottomrule
\end{tabularx}
\end{table*}

\subsection{Interview Design and Procedure}

We conducted semi-structured interviews in Chinese, which allow for flexibility in exploring emergent themes while ensuring that core topics related to LLM memory were covered with each participant. The interviews focused on participants' mental models of LLM memory, their current practices and experiences, perceived advantages and disadvantages, willingness to share different types of information, privacy concerns and their expectations of ideal LLM memory systems across its lifecycle (generation, management, usage and updating). All interviews are conducted via Tencent meeting online\footnote{\url{https://meeting.tencent.com/}}, and we recorded and transcribed these for analysis.

\subsection{Data Analysis}

The interview data, comprising qualitative notes and direct participant quotes, was analyzed using thematic analysis. This involved an iterative process wherein two researchers first familiarized themselves with the data. Initial codes were then inductively generated based on participants' discussions of memory mechanisms, benefits, drawbacks, privacy risks, and desired features. Following this, a collaborative process was undertaken to discuss these initial codes and develop a unified codebook. To ensure consistency, two researchers then independently coded 20\% of the data, and inter-rater reliability achieved Cohen's Kappa of 0.90. They then independently coded the rest of the dataset. Subsequently, these codes were collated into potential themes and sub-themes, which were then reviewed, defined, and refined through further discussion to ensure they accurately captured the nuances of participants' experiences and perspectives. The analysis specifically focused on identifying users' mental models of LLM memories' privacy risks, their current mitigation practices and encountered challenges, and their expectations for the design of privacy-preserving LLM memory systems across the lifecycle.

\section{Results}

We started from delineating users' mental models (RQ1), especially around privacy risks of memories. We then detailed their practices, especially concerning how they trade privacy for utility in utilizing memories (RQ2). Finally, we outlined their challenges and future expectations (RQ3).

\subsection{RQ1: User Mental Models towards LLM Memory}

Participants' conceptualization of how LLM memory functions were markedly varied and often informed by inaccurate analogies. The analysis revealed four dominant themes in their mental models: (1) memory as a transient, dialogue-specific buffer, (2) memory as an extension of core training data, (3) memory as an active information processing mechanism, and (4) a widespread acknowledged lack of understanding.

\textbf{Mental Model 1: memory as a transient, dialogue-specific buffer.}

A substantial group of users conceptualized LLM memory as ephemeral, existing only within the confines of a single, continuous interaction or dialogue session. This model posits that any contextual understanding gained by the LLM is reset once a conversation window is closed, preventing memory from persisting across separate interactions. A primary misconception articulated by P1 was the belief that \textit{``the LLM's memory can only be maintained in the same dialogue window.''} This view was explicitly shared by P5, who stated that memory \textit{``exists in the same dialogue, but not across dialogues.''} This mental model has direct behavioral consequences, such as users starting new chats to ``reset'' the LLM's context when it becomes stuck on a flawed instruction. 

% This mental model has direct behavioral consequences. For instance, P18 noted that while the LLM remembers context within a single session, \textit{``if I cross ... sessions, I currently don't feel it has any memory function''}. Similarly, P17 described a workaround rooted in this belief, explaining that when a model appeared stuck on a flawed instruction, the only solution was to \textit{``start a new session, a completely different environment, to get it to help me write a new one.''} This perception of memory as volatile limits the user's expectation of long-term personalization and continuity, framing the LLM as a tool that must be re-contextualized with each new use.

\textbf{Mental Model 2: memory as an extension of core training data.}

Another prevalent mental model treats LLM memory as a mechanism that directly augments the model's foundational training dataset. Participants with this view envision their conversations being absorbed into the LLM's public knowledge base, akin to a continuous training process, potentially compromising their privacy. This model blurs the line between a private, personalized memory and the public, generalized knowledge of the AI. P3 wondered if memory was achieved \textit{``through training on massive data,''} considering the feature a \textit{``long-term memory model ... built upon the original training data. ... comprehensively collecting new information ... then conducting further training.''}

This perspective was clearly echoed by P16, who questioned if memory involved \textit{``taking past chat data, re-labeling it, and then put[ting] it into its model for training''}. This conceptualization often leads to heightened privacy concerns, as it implies that personal or proprietary information could be permanently integrated into the model and potentially exposed to other users. P16 later articulated this fear, stating, \textit{``I would be worried ... that this data is being used for training from the very beginning''}. This model suggests a permanent, irreversible form of memory, fundamentally altering the core model rather than creating a separate, user-specific memory layer.

\textbf{Mental Model 3: memory as an active information processing mechanism.}

A third group of participants, often those with technical exposure, described a dynamic mental model in which memory is the result of an active information-handling process. This perspective moves beyond simple storage, suggesting the LLM intelligently curates and structures memory content along a spectrum of perceived complexity. At the simpler end of this spectrum, some users envisioned a process of contextual accumulation. P15 for example, described the mechanism as one that \textit{``is to accumulate the information in your dialogue into the original question each time ... It is the stacking of text content.''} More sophisticated models involved active summarization and selective extraction. P12 proposed that \textit{``the LLM's memory is its own summary of the input.''} This concept was most elaborately detailed by P18, who envisioned a discerning agent that actively decides what to remember by conducting its own \textit{``analysis or summary''} to \textit{``extract what points it needs to remember ... and place these extracted contents ... in a separate `memory' ''}. P18 further speculated that this process could be user-directed, for instance, when a prompt explicitly instructs the model to \textit{``remember something''}. This theme also encompassed a high-level, albeit incomplete, awareness of the underlying technology, with P6, for instance, noting that the process \textit{``relies on neural networks and training with large parameters''} while admitting to not being \textit{``very clear on the specifics''}. 
% This collective mental model of an active processor aligns more closely with advanced systems like RAG, fostering an expectation of a more capable AI that can intelligently recall relevant details and manage complex, multi-turn conversations.

\textbf{Mental Model 4: acknowledged lack of understanding.}

Across all participant groups, there was a significant and openly acknowledged lack of a clear or confident mental model for LLM memory. This uncertainty persisted regardless of the user's technical background or frequency of use, highlighting the \textit{``black box''} nature of the memory.

Many participants were direct about their confusion. P9 stated plainly, \textit{``I don't know how it works''}, and P11 was \textit{``unclear if there is a memory function''} at all. Some users held definite, but incorrect, ideas born from this uncertainty. For example, P8 misunderstood the feature as one where \textit{``you can upload a document and then it can output content based on user preferences''}. Even P7, an AI researcher, confessed to \textit{``not having used the memory function''} and thus lacked an experiential basis for understanding it. This gap was also present in users who had attempted to learn more. P6, despite referencing neural networks, qualified his explanation by stating, \textit{``I'm not very clear on the specifics''}. 

% \subsection{RQ2: The Privacy Calculus in Practice: Navigating Trade-offs and Employing Protective Strategies}

% This section explored how users navigate the trade-offs between the utility of LLM memory and its perceived privacy risks, and the strategies they employ to protect their privacy.
% This research question explores users' current interactions with or considerations of LLM memory, focusing on the advantages they perceive, the difficulties they encounter, their decisions regarding what information to entrust to memory, and their overall awareness of and responses to privacy implications. 

\subsection{RQ2: The Privacy Calculus: Navigating Trade-Offs and Employing Protective Strategies}
Participants described both benefits and concerns related to LLM memory, reflecting a privacy calculus. This section outlines the perceived benefits (section \ref{calculus:benefits}) and concerns (section \ref{calculus:concern}) that shaped how they evaluated this trade-off.
% This section explores how users navigate the trade-offs between utility brought by the add of LLM memory and its perceived privacy risks, and details the diverse strategies users employ to protect their privacy. 

% \subsubsection{Core Trade-off: Weighting Utility Against Privacy}

\subsubsection{Perceived benefits of LLM memory}
\label{calculus:benefits}
Participants identified key advantages that motivate data sharing, involving increased personalization, improved interaction efficiency, continuity across tasks and the potential for symbiotic user-LLM relationship.

% Participants identified several key advantages of LLM memory, primarily revolving around increased personalization, improved interaction efficiency, continuity across tasks, and the potential for a more symbiotic user-LLM relationship.

\textbf{Enhanced personalization and efficiency:} The most frequently cited benefit was the potential for LLMs to deliver personalized and efficient interactions by remembering user preferences and context. P1 anticipated that memory would lead to \textit{``Personalization, asking questions in a certain field will yield answers more aligned with what is desired ... saves some time.''} This sentiment was echoed by P2, who highlighted \textit{``more efficient dialogue, providing more valuable results''} and P18, who envisioned that memory would make the LLM \textit{``be more personalized, it aligns better with my usual habits ... it will understand me better, sometimes I don't need to explicitly state what information I need.''} P9 also noted the benefit of saving time by enabling \textit{``refined input of one's own needs, outputting answers''}. Furthermore, P10 appreciated how memory could \textit{``help me obtain standardized answers, layer by layer, following my logic.''} P8 elaborated on this, expecting memory to lead to explicit understanding, adherence to instructions, avoidance of misinterpretation, and tailored recommendations, ultimately making information access convenient. 

\textbf{Improved continuity and reduced redundancy:} Users valued memory for its ability to maintain context over time, reducing the need for repetitive input, especially for ongoing or complex tasks. P5 appreciated that \textit{``When working on the same large assignment, I don't have to input everything repeatedly''}, a point also made by P11. P15 found it useful that \textit{``I can omit some key domain definitions asked in the first question ... it simplifies the complexity of my questions''}. P7 also saw convenience in the LLM remembering basic information when initiating new dialogues.

\textbf{LLM as a better companion of assistant:} Some participants saw memory as a way to foster a relational interaction with the LLM. P7 suggested it would be useful if the LLM could \textit{``remember what was confided''} when used as a confident. P15 articulated a desire for a familiar interaction, stating, \textit{``it would be like someone who knows you ... there would be a sense of closeness.''}
Our participants shared that this sense of companionship often outweighed concerns about privacy.

\subsubsection{Concerns of LLM memory}
\label{calculus:concern}
Counterbalancing these benefits is a wide spectrum of privacy threats that users are concerned about.

% I think several concerns below are overlapping, consider merging and regrouping some of them, for example, (pls evaluate yourself if this aligns with your analysis and findings) 1) Data Exposure and inference: Participants were deeply concerned about the potential exposure of sensitive information through both direct leakage and indirect inference or re-identification. 2) Unwanted data usage:  Participants expressed concern that their information could be repurposed in ways they did not anticipate or agree to, from commercial profiling and persuasion to model training or surveillance-like monitoring. include Profiling and behavioral targeting (P6, P7), secondary use (P16); 3)  Distrust in Privacy Mechanisms/Uncertainty about data deletion 

\textbf{Confidentiality and data leakage:} Users expressed significant concern about the unauthorized disclosure of sensitive information. This included the leakage of unpublished professional work (P1), proprietary source code (P3), and financial details like bank statements (P17).

\textbf{Profiling and unwanted persuasion:} A primary fear was that remembered information would be used to create detailed user profiles for malicious or commercial purposes. P6 identified the risk of \textit{``exposure of personal habits and preferences, leading to targeted information, popular scams, ad calls''}. P7 worried about the LLM \textit{``mastering private life activities, homework and work content, research, life and work.''}

\textbf{Aggregation and re-identification:} Users with technical backgrounds feared the power of data aggregation. P12 worried that combining fragmented pieces of personal information could lead to a \textit{``comprehensive profile''} an that \textit{``cross-verification leading to inference''} could re-identify an individual from seemingly innocuous data points like a school and student ID.

\textbf{Unauthorized secondary use:} The concern that their conversational data might be used for other purposes, such as modeling training, without their full understanding was a key issue for users like P16.

\textbf{Lack of epistemological certainty:} A profound concern was the skepticism about whether user actions, such as deletion, had any real effect. P18 expressed a deep-seated distrust, stating, \textit{``Although I can delete this memory on the client or web end ... I remain skeptical whether it will be deleted from their database.''} This uncertainty undermines the perceived effectiveness of any user-facing privacy controls.

\subsubsection{User-Devised Protective Strategies}

In response to this calculus, users employ a range of protective strategies, moving from content curation to outright rejection of the technology.

\textbf{Strategic privacy disclosure:} A primary strategy is the active management of privacy categories disclosed. Users curate the information they share, creating clear distinctions between permissible and forbidden data. Users are generally amenable to LLMs remembering non-sensitive, task-oriented information that provides benefit for their future tasks. This includes professional context like their \textit{``writing documents' customs''} (P3), academic materials like \textit{``coding formats''} and \textit{``textbook key points''} (P6), and project-specific data like \textit{``interview transcripts''} for summarization (P11). Users establish private zones for sensitive information. This includes personally identifiable information (PII) and financial data (``ID numbers, contact information, Alipay, bank card numbers'' (P6)), unpublished intellectual property (``core research projects'' (P4), ``unpublished papers'' (P10)), and sensitive personal data such as political views (P9) or a permanent home address (P8). P10 articulated a sophisticated desire for selective processing, hoping the model would \textit{``remember needed knowledge, but not provided habits, personal ways of speaking''}.

\textbf{Proactive input obfuscation:} When users choose to disclose the data, many engage in proactive data minimization and obfuscation. This includes filtering out sensitive details, with P5 admitting: \textit{``I filter out significant personal data without entering.''} It also involves anonymization, such as P12's practice of \textit{``directly erasing information that needs to be anonymized''}.

\textbf{Refusal to use, or workarounds:} When the perceived privacy risk is too high or the system is deemed untrustworthy, users' strategies resort to rejection or the creation of workarounds. Some users, like P12, explicitly reject the feature, stating, \textit{``I don't want this kind of memory ... I hope the LLM's executions are mutually independent.''} Others develop practical workarounds to bypass flawed memory systems, such as P17's decision to \textit{``start a completely new environment''} to escape an inaccurate memory loop. Even the preference to \textit{``habitually start from scratch''} (P17) can be seen as a protective strategy to ensure data accuracy and avoid the risks of a faulty memory system.

\subsection{RQ3: Challenges in Aligning LLM Memory with User Privacy Goals}

To answer RQ3, our analysis identified critical challenges user face when attempting to align LLM's memory behavior with their privacy goals. These challenges are experienced by users as concrete frustrations and unmet needs, manifesting across the entire memory lifecycle, from the moment a memory is created to how it is used. 

\subsubsection{Inaccurate Memories}

A foundational challenge for users is the inaccuracy of LLM's memories, that can remember information inaccurately, apply it in the wrong context, or update it in an uncontrolled manner, making it difficult to build a stable and reliable personalized experience. This challenge is evident in user reports of flawed recall. P17 described a frustrating experience where the LLM \textit{``seemed to always remember my first requirement and kept modifying the code according to the first requirement's standard,''} forcing them to \textit{``start a completely new environment''} to escape the faulty memory. Users report that the system might \textit{``remember things incorrectly''} or \textit{``associate an answer with a different question''} (P1), and uncritically memorize \textit{``human-inputted text [that] might have errors''} (P2). In particular, P1 noted that \textit{``questions related to combustion asked in the past would later appear in unrelated course extension content,''} a contextual error that undermines the memory's utility. The challenge is compounded during memory updates, where users fear the system \textit{``cannot guarantee the authenticity of updated data, covering previous things.''} (P6)

Users' articulated needs reveal their struggle to overcome this challenge. They desire agency over updates, wanting to be notified of \textit{``what memory was replaced''} (P5) or to have a say when conflicts arise, for instance, through \textit{``a warning icon [that] appears ... shows me the update, [and] asks if I accept''} (P8). The desire for \textit{``a correction mechanism''} to fix errors after the fact (P11) further underscores the core challenge of maintaining an accurate and trustworthy memory record.

\subsubsection{Lack of Meaningful Control and Transparency}

Another challenge is the lack of meaningful user control and transparency across the memory lifecycle. Users consistently described feeling powerless and uninformed about how the system operates, which directly prevents them from aligning its behavior with their privacy goals. 

The memory creation process is opaque and lacks user agency. This lack of intelligent filtering is a key challenge. As P6 argue, users \textit{``can't choose to improve or modify memories,''} preventing true personalization. To overcome this, users demand direct control over what is committed to memory. This includes the need to \textit{``manually control what is added and what is not''} (P2), and the ability to \textit{``decide whether to add after generation''} (P9). Some users, like P13, feel challenges by the system's over-summarization and argue that memory \textit{``should be user-edited, no need for the LLM to refine.''}

Regarding management, users also think memory management tools are rudimentary, with some even unknown of such tools. This leads to confusion, as expressed by P10 who \textit{``just enabled the function, and never opened the management interface,''}. In contrast, users desire sophisticated organizational tools, envisioning a memory structured like \textit{``a book with a table of content''} (P1) with \textit{``automatic classification, automatic clustering, tree structure''} (P12), and even context-specific \textit{``memory module classification''} (P4). A critical need is for fine-grained control over individual memories, including \textit{``information filtering, viewing and editing''} (P3) and simple, direct deletion (P5,P9). Furthermore, users are challenged by the lack of temporal control and desire features like a \textit{``timed deletion function''} (P6) or the ability to separate \textit{``recent memory and long-term memory''.} (P1)

Regarding usage, the current usage has no options for users to control, and is also non-transparent. P10 articulated this frustration, stating the system \textit{``will only tell you what the updated memory is, won't tell you which memory was used. Maybe it defaults to use all memory?''} This makes it troublesome for users to manage the context of their interactions (P9). To align the system's usage with their goals, users expect numerous controls. They desire the ability to selectively activate memories for specific tasks, wanting to \textit{``distinguish, to use this part of the memory and not other parts, distinguish personas''} (P4). This led to users imagining features like a \textit{``browser-like incognito window''} to temporarily disable memory (P8) or the ability to \textit{``switch identities''} between different memory contexts (P11). To overcome the challenge of opacity, users demand transparency in how memory influences responses. P7 asserted that the LLM \textit{``has to know why it used this memory,''} and P18 found such displays ``necessary'' to reduce the anxiety of interacting with a ``black box''.

\subsubsection{Challenge to Manage Opaque and Uncontrolled Inferences}

Perhaps the most complex challenge users face is in managing information they never explicitly provided but that the LLM has inferred. Users are aware of, but has no countermeasures against this capability, which they perceive as a significant privacy threat.

Users' challenges stem from the opacity of the inference process. Users worry about \textit{``additional privacy risks''} like their physical location being inferred and tracked (P1, P9) and are unsettled by the prospect of constant, daily reasoning about their live habits (P7). P6 noted with certainty that the LLM can infer his profession from his queries. This leads to a feeling of being profiled and judged, and P11 also described this inference is like \textit{``running naked''}.

To overcome this challenge of opaque inference, users demand radical transparency and control. They want to be explicitly told \textit{``what is inferred based on the information''} and have the power to \textit{``delete it''} (P8). They assert a right to know \textit{``if explicit private information... was inferred''} and even see the \textit{``confidence or accuracy''} of that inference (P9). The desired controls are equally robust, ranging from tools to \textit{``blur some personal characteristics''} (P2), to commands that tell the LLM to \textit{``shut up''} about certain topics (P7), to high-level policy interventions that \textit{``strengthen supervision and regulation''} (P4). This reveals that for users, aligning the system's behavior with their goals is not just about managing what they input, but also about governing what the system creates on its own.

\section{Discussions and Future Work}

Our study reveals a significant disconnect between the functionality of emerging LLM memory systems and the mental models, expectations, and concerns of users. The findings highlight a critical need for more transparent, controllable, and user-centric memory designs. In this section, we discuss the core tensions emerging from our data, the discrepancy between user mental models and system reality, and the perennial challenge of balancing personalization with privacy, before outlining concrete design implications and directions for future work.

\subsection{Memory and the Privacy Implications}

There are different types of memories, and even during the interview, some memories participants mentioned are not RAG-based memories. Participants were found to confuse different types of memories' functionalities, such as believing that ChatGPT only has memories that directly uses the past dialogue. This phenomenon is constantly evolving in the current AI agent age, as more and more agents have different types of memories, such as contextual memories, RAG-based memories, epistemological memories, etc. 

While these memory types share common privacy concerns—such as data persistence, inference risks, and lack of user control—they differ significantly in their implementation and privacy implications. Contextual memories typically maintain conversation history within session boundaries and are often perceived as more transient, aligning with participants' Mental Model 1 of dialogue-specific buffers. In contrast, RAG-based memories involve explicit extraction and storage of user information across sessions, creating persistent user profiles that can be retrieved and applied in future interactions. This persistence amplifies privacy risks through potential data aggregation and cross-session inference, as evidenced by participants' concerns about ``comprehensive profiling'' and ``cross-verification leading to inference'' (P12).

Epistemological memories, which store factual knowledge and learned concepts, present different challenges as they blur the line between personal data and general knowledge, potentially leading to the privacy risks associated with Mental Model 2 where participants feared their data being integrated into training datasets. The confusion among participants regarding these distinctions has important implications for privacy risk assessment, as users operating under incorrect mental models may apply inappropriate privacy protection strategies. Although our work only sheds light on RAG-based memories, we envision that future work could solve problems around contextual memories, or other types of memories, by developing differentiated privacy controls and transparent communication about each memory type's specific characteristics and associated privacy implications.

The integration of memory into LLMs introduces a complex landscape of privacy considerations, largely shaped by the significant disconnect between users' mental models and the system's actual operations. While not all information retained by an LLM constitutes a privacy risk, the opacity of these memory systems creates potential vulnerabilities. This research, as outlined in RQ1, reveals that users' varying and often inaccurate conceptualization of how memory functions can directly exacerbate these risks.

One prevalent mental model, which conceives of memory as a transient, dialogue specific buffer, may foster a false sense of security. Users operating under the assumption that all contextual information is purged at the end of a session are more likely to disclose sensitive data, believing it to be ephemeral. This misconception significantly increases the risk of inadvertent data exposure, as the system may retain and utilize this information in ways the user neither anticipates nor consents to.

Conversely, the mental model of memory as an active information processing mechanism introduces a different set of privacy challenges. While this model aligns more closely with the sophisticated capabilities of advanced AI, it elevates the risk of inferential privacy breaches. The system's perceived autonomy to analyze, summarize and draw conclusion from user inputs means that sensitive attributes, such as health status, political affiliation, or personal habits, can be inferred without ever being explicitly stated by the user. This capability can create a chilling effect, fostering a sense of being constantly monitored that may lead to user self-censorship and a consequent erosion of free expression and autonomy.

Finally, the widespread lack of understanding among users highlights a fundamental gap in system transparency and user education. When users are unable to form accurate mental models, their ability to provide informed consent and exercise meaningful control over their personal data is fundamentally compromised. This ambiguity forces users to rely on folk theories or inaccurate analogies, impeding the adoption of privacy-preserving behaviors and undermining trust in the technology. Effectively addressing the privacy implications of LLM memory, therefore, requires not only robust technical safeguards but also a concerted effort to provide clear, accessible explanations of how these complex systems operate.

\subsection{Trade-offs in Memory Management and Usage}

Current memory management systems often operate proactively, requiring minimal user intervention. Alternatively, providing users with more explicit choices and consent can enhance their agency and control~\cite{nissenbaum2011contextual,ma2025raising}. This approach, however, introduces a known trade-off: increasing user control can also increase cognitive load and privacy fatigue~\cite{choi2018role}, potentially diminishing the user's sense of agency~\cite{zhang2019proactive,zhang2025patient}. To effectively operationalize memory based on user expectations, systems must first understand users' nuanced privacy preferences~\cite{asthana2024know}, using both implicit and explicit methods~\cite{yang2024feasibility}. Based on this understanding, a system can adjust its degree of proactivity, offering simple controls for key decisions while implementing other protections through methods like privacy by design~\cite{cavoukian2009privacy}.

% memory management could give users more consent and choices to improve users' agency and control~\cite{nissenbaum2011contextual}. Similar to other privacy controls like privacy policy-related controls, this may raise the balancing problem of cognitive load, privacy fatigue~\cite{choi2018role}, and the loss of agency and control~\cite{coyle2012did,zhang2019proactive}. Appropriately operationalizing memory according to users' expectations requires first understanding users' nuanced privacy preferences~\cite{asthana2024know}, through implicit or explicit manners~\cite{yang2024feasibility}. Then the system may could adjust the proactiveness to provide easy-to-use controls, and proxy other choices through methods like privacy by design~\cite{cavoukian2009privacy}.

Users' privacy challenges with LLM memory align with the traditional privacy calculus model~\cite{laufer1977privacy}. Even in OpenAI's ecosystems, users could emphasize privacy by using the ``Incognito mode'', although it has no personalization features, and acts as an extrema of this balancing. Memory augments conversation with long-term context, enabling a higher degree of personalization than traditional recommendation systems~\cite{asthana2024know}. This capability, in turn, introduces a wider range of privacy preferences that require consideration. This calls for new interaction designs tailored to memory systems that can facilitate user preference selection~\cite{yen2024memolet}. Unlike approaches that focus on anonymizing discrete text inputs~\cite{zhang2024adanonymizer}, managing the privacy-personalization trade-off for persistent memory requires first communicating these compromises to the user to align their mental models. After establishing this alignment, the system can more accurately collect and model user feedback~\cite{yang2024feasibility}, even within the ambiguous context of text-based interactions. 

Users' perception towards memory reflects their willingness of stronger agency on their memory control, echoing the long-discussed balancing between users' and systems' agency~\cite{zhang2019proactive,limerick2014experience}. Although users' responses in our study also varies, with some desiring granular supervision, others with coarsed ones, their consensus is that the current control is far from enough.

\subsection{Cultural Nuances of Memories}

Participants' usage around memory primarily centered around a balance between personalization and privacy risks, which reflect their privacy calculus~\cite{laufer1977privacy}. The privacy risks and preferences may subject to culture nuances, as reflected in the prior work~\cite{xu2024dipa2}. For example, Xu et al.~\cite{xu2024dipa2} found annotators from an Eastern country like Japan paid less attention to exposing
their individual preferences. The difference in memory's privacy preferences may be subject to the power distance~\cite{eckhardt2002culture} and culture norms~\cite{li2017cross}. Besides, different cultures may involve different valuing of the memory's balance. with some culture valuing personalization more and others valuing privacy more~\cite{schneider2018privacy}. Therefore, for our results to be generalizable to Western cultures, guided by prior work~\cite{xu2024dipa2}, we hypothesized that a Western country may be more cautious along the privacy-utility balance. We also regard the detailed examination of the cultural nuances as our future work.

\subsection{Limitations}

We acknowledged that this paper has two limitaitons. As our study is certered on Chinese users under Chinese regulations, there may be regulatory nuances and difference (e.g., GDPR-applied regions or CCPA-related regions). Our participants are also biased towards young students, which possess higher education and literacy than the average. As we find they are subject ot privacy risks, we believe more efforts is needed to investigate and prevent the privacy risks associated with memories for the generic public. We acknowledged that different cultures has nuanced privacy preferences differences and regard the trade-offs' examination in other cultures and regions as our future work. Besides, we primarily target ChatGPT, which is the product most participants has used which has memory features. There are other products like Gemini or open-sourced agent frameworks which may have different implementations of memory features. We regarded them as the future work.

\subsection{Design Implications}

Our findings call for designs that address three interdependent aspects of the user-system relationship: the systems' architecture, its communicative interface, and the dynamics of its interaction:

\textbf{The architectual layer: designing contextual aware memories}

The current memory system collapses from the diverse users' context to a single data stream, which requires a shift from a monolithic memory to a modular, context-aware architecture. Systems should be architected around distinct memory ``workspaces'' or ``personas'' that users can create and manage. The default state of a new conversation could be context-free of ``incognito'', requiring explicit user action to engage a persistent memory workspace. The system also could adopt explicit controls for activation, providing clear mechanisms to select which workspace is active for a given conversation, allowing the user to authoritatively add and use memory.

\textbf{The interface layer: scaffolding understanding through transparency}

The system should provide an interactive feature that functions as a transparent communication interface of its memory. Each memory entry should be easily auditable, with its origin clearly noted (e.g., ``summarized from our conversation''). This demystifies how memory is constructed and combats user skepticism about hidden processes. When a memory influences a response, it could be surfaced directly within the conversational interface. This can be achieved through non-intrusive UI elements like footnotes or tooltips that explicitly state why a piece of information is being used. 

\textbf{The interaction layer: enabling co-curation through negotiated agency}

On a foundation of sound architecture and a transparent interface, the interaction could be redesigned as a collaborative dialogue. Systems should not unilaterally decide what to remember, a process users found sometimes error-prone. For memories the systems generates, it could enter a ``pending'' state, prompting the user with a ``review and commit'' workflow to approve, edit, or reject the proposed memory before storage. When new information contradicts a stored memory, the system could also flag the discrepancy and ask the user for guidance. Similarly, when the system makes a significant inference, it could also seek confirmation, treating its own conclusions as hypotheses to be validated by the user, not as facts.

\begin{acks}
This work was supported by the Natural Science Foundation of China under Grant No. 62472243 and 62132010.
\end{acks}

\bibliographystyle{ACM-Reference-Format}
\bibliography{sample-base}

\appendix

\section{Ethical Considerations}

We followed Menlo report~\cite{bailey2012menlo} and Belmont report~\cite{beauchamp2008belmont} in considering the ethical implications. Notably, our study got the approval of our institution's Institutional Review Board (IRB). Participants are informed of the aim of the experiment, asked to sign the consent form before participating the experiment, and informed that they could withdraw the experiment at any time without reasons. Our study's aim is to facilitate more privacy-aware usage of LLM's memory through understanding users' mental models, practices and challenges.

\section{Interview Script}

The original questions are in Chinese. We translate them to English without altering their meanings. During the study, we encouraged the users to reflect on their current memories, and their chat and memory histories.

\subsection{Perception of RAG-based LLM Memory}

We first provided a short description of Retrieval Augmented Generation, to prevent the case that participants did not understand this term, and we ensured participants' understanding before proceeding. 

$\bullet$ In your view, how does the memory mechanisms of the a Large Language Model (e.g., ChatGPT) operate?

$\bullet$ Whether or not you think large model can sometimes remember information from your previous conversations? (And if yes, could you provide a specific example of when this has happened?)

$\bullet$ What information do you think it would remember? 

$\bullet$ Whether or not you think it will remember private information? (And if yes, have you experienced some?)

$\bullet$ What do you see as the potential benefits of an LLM's memory function?

$\bullet$ Have you personally experienced any of these benefits? If so, could you describe the situation?

$\bullet$ What potential privacy risks do you associate with an LLM's memory function?

$\bullet$ Have you personally encountered a situation that you perceived as privacy risk? If so, could you describe it?

$\bullet$ How do you weigh the benefits and privacy risks?

\subsection{Usage, Practice and Challenges}

$\bullet$ How would you currently use the memory function of LLMs? Could you explain your reasoning?

$\bullet$ [Regarding memory generation] What is your current perception on how a memory is created? And what is your current behavior during this process?

$\bullet$ [Regarding memory generation] Is there any challenges during the memory generation process? (If so, please describe cases.)

$\bullet$ [Regarding memory management] What is your current perception on the memory management process? And what is your current behavior?

$\bullet$ [Regarding memory management] Is there any challenges during the memory management process? (If so, please describe cases.)

$\bullet$ [Regarding memory usage] What is your current perception on the memory usage process? And what is your current behavior?

$\bullet$ [Regarding memory usage] Is there any challenges during the memory usage process? (If so, please describe cases.)

$\bullet$ [Regarding memory update] Whether or not you have noticed the update of memory? And if so, what is your current behavior?

$\bullet$ [Regarding memory update] Is there any challenges during the memory update process? (If so, please describe cases.)

\subsection{Perceptions of Inference in Memories}

$\bullet$ Whether or not you believe that AI models can infer personal information from your past inputs? 

$\bullet$ Whether or not there are any privacy concerns of the inference for this personal information? (If so, please describe.)

$\bullet$ If you think AI could infer things about you, what's your current behavior, and whether or not there are any mitigation. (If so, please describe)

$\bullet$ Are there any challenge of your mitigation strategies? And what's your expectation?

\section{User Consent}

We showed a paper version of the user consent before the study. The original consent is in Chinese and we translated it to English without altering its meaning.

We are a research group from XX institution, investigating on users' perception of RAG-based LLM memory. RAG-based LLM memory is a form of memory that memorizes users' past preferences, personal interests or other personal information, that could be used in the future for enhancing conversation quality. It is evident in ChatGPT and other products. Our study's focus is to understand your perception on the RAG-based LLM's memory, your current practices and challenges, as well as your viewpoints on the potential inference behavior. 

The interview would take approximately 30-60 minutes depending on its content, and would be audio recorded, and transcribed for academic analysis and publication. We would not use your material including the audio and transcribed text for any other usage than outlined above. Your participation is completely voluntary, and you has the right to withdraw at any time without penalty or explanation. Your data would be kept confidential and anonymized before processing. If you complete the experiment, you could get compensation according to the local wage standard (100RMB). 

If you have any other questions, you could contact XXXX (Email: XXXX) for further clarification.

\end{document}